\documentclass[epj,referee]{svjour}
\usepackage{graphics}
\begin{document}
\title{QED Corrections to the Electromagnetic Abraham Force}
\subtitle{Casimir Momentum of the Hydrogen atom?}
\author{B.A. van Tiggelen\inst{1} \and S. Kawka \inst{1,2} \and G.L.J.A. Rikken \inst{3}
}                     
\offprints{}          
\institute{ Universit\'e Grenoble 1/CNRS, LPMMC, Maison des
Magist\`{e}res, 38042 Grenoble, France  \and Scuola Normale
Superiore, Physics Department, Piazza dei Cavalieri, 7 56126 Pisa,
Italy \and Laboratoire National des Champs Magn\'etiques Intenses,
UPR3228 CNRS/INSA/UJF/UPS, Toulouse \& Grenoble, France }
\date{Received: date / Revised version: date}
%
\abstract{ We develop a QED approach to find the contribution of the
quantum vacuum to the electromagnetic Abraham force. Semi-classical
theories predict diverging contributions from the quantum vacuum. We
show that the divergencies disappear by Kramers-Bethe
mass-renormalization. The finite remainder is compared to the
relativistic corrections to the Abraham force. This work generalizes
an earlier paper \cite{kawka}, dedicated to the harmonic oscillator,
to the hydrogen atom and \emph{corrects} two subtle errors.
} 
\maketitle
\section{Introduction}
\label{intro} The Abraham force refers to the small force exerted by
time-dependent electromagnetic fields on neutral matter. Despite our
 complete knowledge on classical electromagnetism, a longstanding
 controversy exists about the precise expression of this force, given
 macroscopic, phenomenological constants such as dielectric constant
 and magnetic permeability. This so-called ``Abraham-Minkowski" controversy stems
 mainly from the fact that the macroscopic Maxwell's equations seem to favor a
 radiative momentum density $\mathbf{D }\times \mathbf{B}$
 \cite{jackson} (the Minkowski version), whereas the microscopic version leads to
  $\varepsilon_0\mathbf{E }\times
 \mathbf{B}$ (the Nelson version \cite{nelson}). In addition, neither one of them is in general
 equal to $\varepsilon_0\mu_0\mathbf{E }\times \mathbf{H}$, that is radiative momentum
 does not always seem to be just equal to energy flow $\mathbf{S}=c_0^{-1}\mathbf{E }\times \mathbf{H}$, divided by $c_0$ (the Abraham version, also advocated by Jackson \cite{jackson}).
 The different versions lead to different expressions for the
 Abraham force \cite{brevik}.

Following Nelson \cite{nelson}, it is instructive to calculate
directly the force density $\mathbf{f}= \partial_t (\rho_m
\mathbf{v})$ exerted on the matter, given by the Lorentz force
$\mathbf{f}=\rho_q \mathbf{E} + \mathbf{J}_q  \times\mathbf{ B}$
with $\rho_q = \nabla\cdot \mathbf{P}_q$ and
$\mathbf{J}_q=\nabla\times \mathbf{M}_q -\partial_t\mathbf{ P}_q$
the bound charge density and bound charge current density. For
nonmagnetic materials without free charges this leads to

\begin{eqnarray}\label{cons}
  \partial_t \left(\rho_m \mathbf{v}-\mathbf{P}_q   \times \mathbf{B}\right) &=& \nabla\cdot \mathbf{U} -
  \frac{1}{2}(\nabla\varepsilon_r)\, \varepsilon_0\mathbf{E}^2
\end{eqnarray}
with $\mathbf{U}$ a second rank stress tensor specified elsewhere
\cite{reply}. Since the space integral of $\nabla\cdot \mathbf{U} $
vanishes, and so does the last term in a sufficiently homogeneous
field, the force on the atom is equal to  the classical ``Abraham
force",

\begin{equation}\label{abraham}
    \mathbf{F}_A= \partial_t \int d\mathbf{r} (\mathbf{P}_q
  \times \mathbf{B})= \varepsilon_0 (\varepsilon_r -1) V \partial_t (\mathbf{E}
  \times \mathbf{B})
\end{equation}
 Only a few observational claims exist, such as the work by Walker
\cite{walker}, and our recent work on gases \cite{geertprl12}.

Since a few years several authors have discussed the possibility of
a contribution to the Abraham force stemming from the
electromagnetic quantum vacuum. In the remainder we will refer to
the momentum associated with this QED contribution to the Abraham
force as ``Casimir momentum". Its existence seems even more
controversial. The first approaches are based on a semi-classical
analysis, i.e. classical for the matter, and quantum for the
radiation. Following the early work in Ref.~\cite{feigel}, Refs.
\cite{epj} and \cite{brevik2} demonstrate that Casimir momentum can
exist in so-called bi-anisotropic media. These are media where
electric fields induce magnetic moments, and by symmetry, magnetic
fields induce electric moments. All media can be made
bi-anisotropic, for instance by moving them or by exposing them to
crossed electric and magnetic fields. However, two major problems
show up for Casimir momentum in such media. First, the end result
diverges significantly in the UV and needs to be regularized, and
the question is whether the effect survives the regularization.
Secondly, the question about Lorentz-invariance arises, and more
generally, the need to include relativistic effects. In a recent
Reply \cite{reply} a fully Lorentz-invariant model was discussed in
which Casimir momentum exists, although still with a diverging
value.

In Ref.~\cite{feigel} the divergent frequency integral was solved by
a simple cut-off frequency in the X-ray regime, as had previously
been proposed by Schwinger \cite{schwinger} to calculate the Casimir
energy of an oscillating bubble. Dimensional regularization was
proposed by Brevik etal \cite{miltonbrevik}. When applied to Casimir
momentum \cite{epj}, one finds values inversely proportional to the
sample size but still resulting in extremely small values for
micro-sized media. In a recent work \cite{kawka} we considered the
quantum-mechanical harmonic oscillator exposed to quasi-static
fields as well as to the quantum vacuum.  We showed that Kramer's
mass renormalization \cite{bethe,milloni} eliminates the
divergencies. Unfortunately, this work suffers from two errors that
will be repaired in this work. In addition we shall consider the
hydrogen atom with Coulomb interaction. This will allow us to
conclude that relativistic effects are   (here) much smaller,
typically by a factor $m_e/M$.

\section{UV catastrophe in bi-anisotropic media}
\label{sec:1} The problem that occurs in bianisotropic media is most
easily illustrated by considering an infinite, dispersionless
dielectric medium moving at speed $\mathbf{v}$ with respect to the
observer. Such medium is subject to the Fizeau effect and has a
bi-anisotropic coupling tensor equal to
$g_{ij}=\varepsilon_0(1-\varepsilon_r)\epsilon_{ikj } v_k$, i.e the
magnetic field induces a polarization density equal to
$\varepsilon_0(1-\varepsilon_r)\mathbf{v}\times \mathbf{B}$. It is
straightforward to see that the quantum expectation value of the
Abraham momentum density $\mathbf{P}_\mathrm{A}=-\mathbf{P}_q\times
\mathbf{B}$ is given by $(2\mathbf{v}/3c_0^2) (\varepsilon_r-1)
\langle \mathbf{B}^2/\mu_0\rangle$ and is directed along the
direction $\mathbf{v}$.  Defining the "Casimir inertial mass
density" $\rho_C$ as $\mathbf{P}_A=\rho_C \mathbf{v}$ that in
Eq.~(\ref{cons}) adds up to the momentum density $\rho_m \mathbf{v}$
of ``matter" , gives
\begin{equation}\label{massuv}
    \rho_C = \frac{2}{3}\frac{\hbar}{\pi^3c_0^5}\int_0^\infty {d\omega} (\varepsilon_r-1) \omega^3
\end{equation}
This integral diverges. Usually the divergence of Casimir energy
poses no problem as long since the observable forces are finite. The
Casimir energy between two ideal mirrors is known to have a finite
inertial mass \cite{jaekcel}. Being an observable quantity itself,
the diverging inertial mass density $\rho_C$ does pose a problem. We
can identify $\langle \mathbf{E}\cdot \mathbf{P}_q \rangle = \int
d\omega (\varepsilon_r -1) \hbar\omega^3/\pi^3c_0^3$ as the
potential energy density of the quantum vacuum. Indeed, what we
might expect to find for $\rho_C$ is (the inertial mass due to) the
binding energy associated with Coulomb, Van-der-Waals and Casimir
Polder forces \cite{polder} that can be viewed as longitudinal and
transverse degrees of freedom of the vacuum electromagnetic field,
and which are also known to find their way to the dielectric
constant \cite{pr}. Apart from the divergence, the front factor
$2/3$ in Eq.(\ref{massuv}) is also strange, and reminiscent of the
problem of electromagnetic self-mass of the electron \cite{jackson}.

If we assume free electron optical dispersion $(\varepsilon_r=1-n_e
e^2/\varepsilon_0 m_e\omega^2)$ at high frequencies \cite{jackson},
the divergence of Casimir momentum will still persist as $\int
d\omega \omega$. If we assume that frequencies larger than $\pi/r_e$
do not contribute by some unknown physical principle (valid at
length scales smaller than the free electron radius) the mass
density (\ref{massuv}) would typically be equal to $n_e m_e/\alpha$,
i.e. a factor $1/\alpha$ larger than the actual mass density of the
electrons. It seems likely that this mass is already counted in the
physical values attributed to the masses of particles, as was
already suggested in literature \cite{milloni}. In this paper we
shall validate this argument explicitly for the hydrogen atom, both
for the Fizeau effect and for the magneto-electric effect. For the
first this is quite naturally suggested by Eq. (\ref{massuv}). For
magneto-electric materials the quantum vacuum gives a similar UV
divergence \cite{feigel} for which mass renormalization is much less
obvious.

In the following we consider the quantum mechanics of a hydrogen
atom exposed to crossed, quasi-static electromagnetic fields. We
calculate the total (pseudo-) momentum of the atoms and identify
diverging terms when coupled to the quantum vacuum. We regularize
the infinities, and finally find the finite contribution of the
quantum vacuum to the momentum, into lowest order of the fine
structure constant $\alpha$.

\section{A moving hydrogen atom coupled to EM quantum vacuum and external fields}

We consider a hydrogen atom moving at non-relativistic speed, in
crossed electric and magnetic fields $\mathbf{E}_0$ and
$\mathbf{B}_0$ respectively. Particle 1 is the proton, particle 2
the electron. We shall use the reduced coordinates $\mathbf{R}=(m_1
\mathbf{r}_1+ m_2\mathbf{r}_2)/M$ and
$\mathbf{r}=\mathbf{r}_1-\mathbf{r}_2$ for the center of mass
position and the interparticle distance, with \emph{conjugate}
momenta $\mathbf{P}=\mathbf{p}_1+\mathbf{p}_2$ and $\mathbf{p}=\mu
(\mathbf{p}_1/m_1 -\mathbf{p}_2/m_2)$. The presence of an external
stationary magnetic field $\mathbf{B}_0$ will be treated in the
Coulomb gauge and it is instructive to make the unitary
transformation $U=\exp[ie (\mathbf{B}_0\times \mathbf{R})\cdot
\mathbf{r}/\hbar]$ that conveniently removes $\mathbf{R}$ from the
Hamiltonian \cite{diddel}. We shall write
$H=H_0+H_{\mathrm{rel}}+H_F+W$ and denote with $\tilde{H}=U^*HU $
the transformed Hamiltonian. The nonrelativistic, transformed
Hamiltonian of the atom is given by

\begin{eqnarray}\label{h0}
   \widetilde{H}_0&=&\frac{1}{2\mu}\left(\mathbf{p}+\frac{\Delta m}{M} \frac{e}{2}\mathbf{B}_0 \times \mathbf{r}
   \right)^2 +  \frac{1}{2M}\left(\mathbf{P}  - {e}\mathbf{B}_0 \times
   \mathbf{r} \right)^2 \nonumber \\&&-\frac{e^2}{4\pi \varepsilon_0 r} - e\mathbf{E}_0\cdot \mathbf{r}
\end{eqnarray}
with $\Delta m=m_1 - m_2>0$ \cite{kawka}. This transformed
Hamiltonian commutes clearly and conveniently with the total
conjugated momentum $\mathbf{P}$, even when $\mathbf{E}_0$ depends
on time, meaning that (when transformed back) the pseudo momentum
$\mathbf{Q}=U^*\mathbf{P}U=\mathbf{P}+\frac{e}{2}\mathbf{B}_0 \times
\mathbf{r}=\mathbf{P}_{\mathrm{kin}}+e\mathbf{B}_0 \times
\mathbf{r}$ is a conserved quantity when time-evolution is governed
by $H_0$, as in the classical theory. The relativistic correction
$H_{\mathrm{rel}}$ will be added later. The transverse degrees of
freedom of the electromagnetic quantum vacuum are described by the
Hamiltonian
\begin{equation}\label{hf}
    H_F = \sum_{\mathbf{k}{\epsilon}} \hbar \omega_{k}
\left[ a_{\mathbf{k} {\epsilon}}^\dag a_{\mathbf{k}{\epsilon}} + \frac{1}{2} \right]\\
\end{equation}
The interaction with the quantum vacuum, when transformed reads
\cite{kawka},
\begin{eqnarray}\label{w}
  \tilde{W} &=& -\frac{e}{m_1}\left(\mathbf{p}+\frac{m_1}{M}\mathbf{P}-\frac{e}{2}\mathbf{B}_0\times \mathbf{r}\right)\cdot
  \mathbf{A}\left(\mathbf{R}+
  \frac{m_2}{M}\mathbf{r}\right) \nonumber \\
  && -\frac{e}{m_2}\left(\mathbf{p}-\frac{m_2}{M}\mathbf{P}+\frac{e}{2}\mathbf{B}_0\times
\mathbf{r}\right)\cdot
\mathbf{A}\left(\mathbf{R}-\frac{m_1}{M}\mathbf{r} \right)
  \nonumber \\
  &\, &
+\frac{e^2}{2m_1}\mathbf{A}^2\left(\mathbf{R}+
\frac{m_2}{M}\mathbf{r}\right) +
\frac{e^2}{2m_2}\mathbf{A}^2\left(\mathbf{R}-
\frac{m_1}{M}\mathbf{r}\right)
\end{eqnarray}
where the vector potential of the electromagnetic field is given by
\begin{equation}\label{A}
    \mathbf{A}(\mathbf{x})= \sum_{\mathbf{k\epsilon}} \mathcal{A}_\mathbf{k} \hat{\epsilon} \left[
    a_{\mathbf{k\epsilon}}\exp(i\mathbf{k\cdot x})+a_{\mathbf{k\epsilon}}^\dag\exp(-i\mathbf{k\cdot
    x})\right]
\end{equation}
with $\mathcal{ A}_{k}= \sqrt{\hbar/2\varepsilon_0 c_0 k V}$ in a
quantization volume $V$. In the presence of the quantum vacuum the
pseudo momentum that commutes with $\tilde{H}$ is
\begin{equation}\label{K}
   \mathbf{\tilde{K}}=\mathbf{P}+e \Delta \mathbf{A} + \sum_{\mathbf{k}{\epsilon}} \hbar
    \mathbf{k} a_{\mathbf{k} {\epsilon}}^\dag
    a_{\mathbf{k}{\epsilon}}
\end{equation}
Because $\mathbf{K}$ is a conserved quantity, the time variation of
$\mathbf{P}_{\mathrm{kin}}$, and thus the total exerted force,
equals minus the time-derivative of all other contributions to
$\mathbf{K}$. If $|\Psi_0\rangle$ is the ground state of the joint
atom and photon field, with $|\tilde{\Psi}_0\rangle =
U^*|\Psi_0\rangle$ the transformed ground state, the pseudo-momentum
follows from,
\begin{eqnarray}\label{KKK}
    \langle \mathbf{K}\rangle &=&\langle  \tilde{\Psi}_0 |
    \tilde{\mathbf{K}}| \tilde{\Psi}_0\rangle \nonumber \\ &=& \langle  \tilde{\Psi}_0 |
    \tilde{\mathbf{P}}_{\mathrm{kin}}| \tilde{\Psi}_0\rangle +e^2\mathbf{B}_0 \times \langle  \tilde{\Psi}_0 |
    \mathbf{r}| \tilde{\Psi}_0\rangle \nonumber \\ && + e \langle  \tilde{\Psi}_0 |
    \Delta \mathbf{A} | \tilde{\Psi}_0\rangle + \langle  \tilde{\Psi}_0 |\sum_{\mathbf{k}{\epsilon}} \hbar
    \mathbf{k} \, a_{\mathbf{k} {\epsilon}}^\dag
    a_{\mathbf{k}{\epsilon}}
    | \tilde{\Psi}_0\rangle \nonumber \\
    \, &\equiv&\langle\mathbf{P}_{\mathrm{kin}}\rangle +
    \mathbf{P}_{\mathrm{A}}
    +\delta\langle\mathbf{P}_{\mathrm{pola}}\rangle
    +\langle\mathbf{P}_{\mathrm{long}}\rangle +\langle\mathbf{P}_{\mathrm{trans}}\rangle
\end{eqnarray}
The last two terms are the ``longitudinal" and ``transverse"
contributions to the momentum stemming from the electromagnetic
quantum vacuum \cite{cohen}. The classical Abraham momentum
$\mathbf{P}_{\mathrm{A}}$ is found from the second term in on the
righthand side when we ignore the coupling with the quantum vacuum.
In this case $-e\mathbf{E_0\cdot r}$ becomes the only perturbing
potential and we find,
\begin{eqnarray}\label{abquantum}
| \tilde{\Psi}_0(W=0)\rangle &= &\left[ | 0\rangle + \sum_{j\neq
0}|j\rangle\frac{\langle j| e\mathbf{E_0\cdot r}| 0\rangle}{E_j-E_0}
\right]\otimes | \{0\} \rangle \nonumber \\ &\equiv&
|0,\mathbf{E}_0\rangle \otimes | \{0\} \rangle
\end{eqnarray}
where $| \{0\} \rangle$ denotes the quantum vacuum, empty for all
plane waves, and $|j \rangle$ represent the unperturbed eigen states
of the atom. We will denote by $|0,\mathbf{E}_0\rangle$ the atomic
ground state with explicit reference to the the external electric
field, and given by the first factor in the above expression.
Insertion directly generates $ \mathbf{P}_{\mathrm{A}} =
\varepsilon_0\alpha(0) \mathbf{B}_0 \times \mathbf{E}_0$ with the
well-known expression for the static polarizability,
\begin{equation}\label{alpha0}
    \alpha(0) = \frac{2}{3}\frac{e^2}{\varepsilon_0}\sum_{j\neq 0} \frac{|\langle j| \mathbf{r}|0\rangle|^2 }{E_j-E_0}
\end{equation}
which for the hydrogen atom equals $\alpha(0)=18\pi a_0^3$
\cite{feynmann}.

To find the contribution of the quantum vacuum to $\mathbf{K}$, we
need to know how the ground state $|i=0,\mathbf{Q}_0,\{n=0\}\rangle$
of an atom with pseudo-momentum $\mathbf{Q}_0$ in an empty vacuum is
perturbed by the coupling $W$. Let $| j \mathbf{Q} \{n\} \rangle$ be
an eigen state of $\tilde{H}_0$, $\mathbf{P}$ and $\tilde{H}_F$, the
second with eigenvalue $\mathbf{Q}$. We will restrict ourself to
processes involving absorption and creation of one photon, i.e.
involving the products of only two $\mathbf{A}(\mathbf{r})$ photon
fields. The ground state is, up to second order in the coupling,
equal to (the accent $'$ is to avoid summing over ground state
itself),

\begin{eqnarray}
  |\tilde{\Psi}_0 \rangle &=& \left[1- \frac{1}{2} \sum_{i\mathbf{Q}n} '\frac{|W_{i\mathbf{Q}n,0\mathbf{Q}_00 }|^2}{(E_{0\mathbf{Q}_00}-
  E_{i\mathbf{Q}n})^2}\right]| 0 \mathbf{Q}_0 \{0\} \rangle \nonumber \\ &+ &\sum_{i\mathbf{Q}n}  '
  \frac{W_{i\mathbf{Q}n,0\mathbf{Q}_00}}{E_{0\mathbf{Q}_00}- E_{i\mathbf{Q}n}}
  |i\mathbf{Q}n\rangle  \nonumber \\
  \,  &+&
 \sum_{i\mathbf{Q}n} ' \sum_{i'\mathbf{Q}'n'}  '
  \frac{W_{i\mathbf{Q}n,i'\mathbf{Q}'n' }W_{i'\mathbf{Q}'n',0\mathbf{Q}_00 }}{(E_{0\mathbf{Q}_00}- E_{i\mathbf{Q}n})
  (E_{0\mathbf{Q}_00}- E_{i'\mathbf{Q}'n'})} |i\mathbf{Q}n\rangle \nonumber \\& -& W_{0\mathbf{Q}_00,0\mathbf{Q}_00
  }  \sum_{i\mathbf{Q}n} '\frac{W_{i\mathbf{Q}n,0\mathbf{Q}_00 }}{(E_{0\mathbf{Q}_00}-
  E_{i\mathbf{Q}n})^2} \rangle|i\mathbf{Q}n\rangle
\end{eqnarray}

Note that the $\mathbf{A}^2(\mathbf{r})$ terms contained in $W$ do
not contribute in this expression since $\langle i\neq 0
|\mathbf{A}^2(\mathbf{r}) | 0\rangle =0$ for any atomic state other
than the ground state. Note also that the last term in the perturbed
wave function does not contain one-photon processes since $\langle
\{0\} |\mathbf{A}(\mathbf{r}) | \{0\}\rangle =0$. Several terms
yield diverging contributions to the momentum $\mathbf{K}$ and will
below be identified. The factor in the first term proportional to
the unperturbed ground state is imposed by normalization. In
Appendix A we estimate it to be of order $\alpha^3$, which makes it
beyond the scope of this work.

\subsection{$\delta\langle\mathbf{P}_{\mathrm{pola}}\rangle $}

The momentum $\delta\langle\mathbf{P}_{\mathrm{pola}}\rangle $ can
be regarded as a quantum vacuum contribution to the static
polarizability at $\mathbf{B}_0=0$ that finds its way to the Abraham
momentum via the classical expression. As such it is arguably not a
true ``Casimir" momentum. It takes the form,

\begin{eqnarray}
  &&\delta\langle\mathbf{P}_{\mathrm{pola}}\rangle = e\mathbf{B}_0 \times \langle 0,\mathbf{Q}_0,\mathbf{E}_0  \{0\}|
  \mathbf{r} \sum_{i\mathbf{Q}n} ' \sum_{i'\mathbf{Q}'n'}  '
  \nonumber \\
  && \frac{W_{i\mathbf{Q}n,i'\mathbf{Q}'n' }W_{i'\mathbf{Q}'n',0\mathbf{Q}_00 }}{(E_{0\mathbf{Q}_00}- E_{i\mathbf{Q}n})
  (E_{0\mathbf{Q}_00}- E_{i'\mathbf{Q}'n'})} |i\mathbf{Q}n \rangle + c.c.
\end{eqnarray}
The matrix elements of $W$ contain exponentials of the form
$\exp(\pm i\mathbf{k}\cdot \mathbf{r}_i)$ stemming from
$\mathbf{A}(\mathbf{r})$. Those terms for which products occur of
 exponentials with opposing phase,  diverge. They can be
 re-arranged to give,

\begin{eqnarray}
  &&\delta\langle\mathbf{P}_{\mathrm{pola}}\rangle_{\mathrm{div}} =  2\, \mathrm{Re} \frac{e^2}{m_1^2}\sum_{\mathbf{k\epsilon}}
  \mathcal{A}_k^2 \nonumber \\
 && \langle 0,\mathbf{E}_0| e \left(\mathbf{B}_0 \times \mathbf{r} \right) \frac{1'}{\tilde{H}_0(\mathbf{p},\mathbf{Q}_0)-E_{0\mathbf{Q}_00}}
 \left(\mathbf{p}+\frac{m_1}{M}\mathbf{Q}_0 \right)\cdot \mathbf{\epsilon}\nonumber\\
 &&  \frac{1}{\tilde{H}_0(\mathbf{p}+\frac{m_2}{M}\hbar\mathbf{k},\mathbf{Q}_0-\hbar\mathbf{k})-
  E_{0\mathbf{Q}_00}+\hbar \omega_k}\left(\mathbf{p}+\frac{m_1}{M}\mathbf{Q}_0\right)\cdot
  \mathbf{\epsilon} |0,\mathbf{E}_0\rangle   \nonumber
\end{eqnarray}
plus a similar term for $m_2$. The extra momentum and energy in
denominator of the second line stem from the absorption of a virtual
photon in the intermediate photon state $\{n\}$. This changes the
total state and thus removes the accent in the sum over the
remaining atomic states. It generates a term $\hbar^2k^2/2m_{1(2)} +
\hbar \omega_k$ and leads to a
 $k$-integral that diverges in the UV. Upon extracting the divergency leads
 to,
 \begin{eqnarray*}
   && \delta\langle\mathbf{P}_{\mathrm{pola}}\rangle_{\mathrm{div}} = \nonumber \\ && \frac{1}{2}\frac{\delta m_1}{m_1^2}
    \langle 0,\mathbf{E}_0| e \left(\mathbf{B}_0 \times \mathbf{r} \right) \frac{1'}{\tilde{H}_0-E_{0\mathbf{Q}_00}}
 \left(\mathbf{p}+\frac{m_1}{M}\mathbf{Q}_0 \right)^2 |0,\mathbf{E}_0\rangle \nonumber \\ && + \frac{1}{2}\frac{\delta m_2}{m_2^2}
    \langle 0| e \left(\mathbf{B}_0 \times \mathbf{r} \right) \frac{1'}{\tilde{H}_0-E_{0\mathbf{Q}_00}}
 \left(\mathbf{p}-\frac{m_2}{M}\mathbf{Q}_0 \right)^2
 |0,\mathbf{E}_0\rangle \nonumber \\
&& +c.c \nonumber
 \end{eqnarray*}
where we have introduced the masses,

\begin{equation}\label{qedmass}
    \delta m_i=\frac{4}{3\pi }\alpha \hbar^2 \int_0^\infty \, \frac{dk k }{\hbar^2k^2/2m_i + \hbar\omega_k}
\end{equation}
The diverging mass is recognized as the nonrelativistic QED
contribution to the bare mass \cite{milloni}, and is supposed to be
absorbed by the value attributed to the observed mass. This
''Kramer-Bethe" mass renormalization method seems reasonable
provided that all masses acquire the same diverging contribution.
The polarizability is in principle a complicated function of the
reduced mass. It is easy to check that $\delta(1/\mu) =-{\delta
m_1}/{m_1^2}- {\delta m_2}/{m_2^2}$. As a result, the $\mathbf{p}^2$
term in the expression for
$\delta\langle\mathbf{P}_{\mathrm{pola}}\rangle_{\mathrm{div}} $
contributes
\begin{eqnarray*}
   &&\delta\langle\mathbf{P}_{\mathrm{pola}}\rangle_{\mathrm{div}}
   =\nonumber \\&&
 e\mathbf{B}_0 \times  \langle 0,\mathbf{E}_0| \mathbf{r} \frac{-1'}{\tilde{H}_0 -E_0}
\delta(1/2\mu)\mathbf{p}^2 |0,\mathbf{E}_0\rangle + c.c \nonumber
\\ &&= e\mathbf{B}_0 \times \left( \langle 0,\mathbf{E}_0| \
\mathbf{r} |\delta 0,\mathbf{E}_0\rangle
+ c.c. \right) \nonumber \\
&&= \varepsilon_0 \delta \alpha(0)\mathbf{B}_0 \times  \mathbf{E}_0
\end{eqnarray*}
where $ |\delta 0,\mathbf{E}_0\rangle = -(\tilde{H}_0-E_0)'^{-1}
\delta \tilde{V}| 0,\mathbf{E}_0\rangle$ is the first-order
modification of the atomic ground state due to the perturbation
$\delta \tilde{V} =\delta(1/2\mu)\mathbf{p}^2 $ in the kinetic
energy.  We thus conclude that this divergence disappears when all
masses in $\alpha(0)$   absorb  consistently the same diverging QED
contribution~(\ref{qedmass}).

The terms proportional to $\mathbf{Q}_0^2$ and $\mathbf{p}\cdot
\mathbf{Q}_0$ in $
\delta\langle\mathbf{P}_{\mathrm{pola}}\rangle_{\mathrm{div}}$ can
be seen to vanish.  The other terms in
$\delta\langle\mathbf{P}_{\mathrm{pola}}\rangle$ are finite. A
straightforward analysis comparable to the one for
$\delta\langle\mathbf{P}_{\mathrm{long}}\rangle$ below shows it be a
factor $\alpha^2m_e/M$ smaller than $\mathbf{P}_A$. In this work we
shall restrict ourselves to orders $\alpha^2$. We conclude that no
quantum vacuum contribution of the order $\alpha^2$ exists to the
static polarizability.

\subsection{$ \langle\mathbf{P}_{\mathrm{long}}\rangle $}

The quantum expectation of the momentum of the longitudinal vacuum
field reads,

\begin{eqnarray}
  \langle\mathbf{P}_{\mathrm{long}}\rangle &=&
   \langle 0 ,\mathbf{E}_0, \mathbf{Q}_0, \{0\}|
  \left(e\Delta \mathbf{A}(\mathbf{r}_1) - e\Delta \mathbf{A}(\mathbf{r}_2)\right)\times\nonumber \\  &&\sum_{i\mathbf{Q}n} '
  \mathcal{A}_k^2
  \frac{W_{i\mathbf{Q}n,0\mathbf{Q}_00 }}{(E_{0\mathbf{Q}_00}- E_{i\mathbf{Q}n})
  } |i\mathbf{Q}n \rangle + c.c.
\end{eqnarray}

If we acknowledge the subsequent annihilation and creation of one
virtual photon, we find that
\begin{eqnarray}\label{long}
 &&\langle \mathbf{P}_{\mathrm{long}}\rangle =  2e^2\mathrm{Re}  \sum_{\mathbf{k\epsilon}}  \mathcal{A}_k^2  \nonumber \\
  && \left\{  \langle 0,\mathbf{E}_0 |
 \left( 1 -  e^{-i\mathbf{k}\cdot \mathbf{r}} \right) \times
 \right. \nonumber \\
 &&\frac{1}{\tilde{H}_0(\mathbf{p}-\frac{m_2}{M}\hbar\mathbf{k},\mathbf{Q}_0-\hbar\mathbf{k})-
  E_{0\mathbf{Q}_00}+ \hbar\omega_k} \nonumber \\ && \epsilon\cdot \left(\frac{\mathbf{p}}{m_1} + \frac{\mathbf{Q}_0}{M}-\frac{e}{2m_1}
  \mathbf{B}_0\times \mathbf{r}\right) | 0 ,\mathbf{E}_0\rangle   \nonumber \\
  &+&
  \langle 0,\mathbf{E}_0 |
 \left( e^{i\mathbf{k}\cdot \mathbf{r}} - 1 \right)\times \nonumber \\
 &&\frac{1}{\tilde{H}_0(\mathbf{p}+\frac{m_1}{M}\hbar\mathbf{k},\mathbf{Q}_0-\hbar\mathbf{k})-
  E_{0\mathbf{Q}_00}+ \hbar\omega_k} \nonumber \\ &&\left.\epsilon\cdot \left(\frac{\mathbf{p}}{m_2} - \frac{\mathbf{Q}_0}{M}+\frac{e}{2m_2}
  \mathbf{B}_0\times \mathbf{r}\right) | 0 ,\mathbf{E}_0\rangle
  \right\}\nonumber
\end{eqnarray}
It is clear that terms involving exponents $\exp(\pm
i\mathbf{k}\cdot \mathbf{r})$ ensure finite $k$-integrals. We will
first focuss on the diverging terms. We can extract the divergence
by keeping only recoil plus photon energy in the denominator, to get
\begin{eqnarray}
 \langle\mathbf{P}_{\mathrm{long}}\rangle_{\mathrm{div}} &= &\delta
m_1 \langle 0,\mathbf{E}_0 | \left(\frac{\mathbf{p}}{m_1} +
\frac{\mathbf{Q}_0}{M}-\frac{e}{2m_1}
  \mathbf{B}_0\times \mathbf{r}\right) | 0,\mathbf{E}_0 \rangle \nonumber \\
  &+& \delta m_2 \langle 0,\mathbf{E}_0 |\left(-\frac{\mathbf{p}}{m_2} + \frac{\mathbf{Q}_0}{M}-\frac{e}{2m_2}
  \mathbf{B}_0\times \mathbf{r}\right) | 0,\mathbf{E}_0\rangle \nonumber
\end{eqnarray}
with the masses $\delta m_i$ defined earlier in Eq.~(\ref{qedmass}).
From the relation $[\mathbf{r}, \tilde{H}_0]=i\hbar
\partial_\mathbf{p} \tilde{H}_0$ it is easy to show that $\langle 0,\mathbf{E}_0
| \mathbf{p} + (\Delta m/M) (e/2) \mathbf{B}_0 \times \mathbf{r }|
0,\mathbf{E}_0 \rangle=0$. This leads to
\begin{eqnarray}\label{longfinal}
\langle\mathbf{P}_{\mathrm{long}}\rangle_{\mathrm{div}}
&=&\left(\mathbf{Q}_0 -
 \frac{e}{2} \langle 0,\mathbf{E}_0 | \mathbf{B}_0\times \mathbf{r}| 0 ,\mathbf{E}_0\rangle\right) \frac{\delta m_1+\delta m_2}{M}
 \nonumber \\ &=& \langle 0,\mathbf{E}_0 | \mathbf{\tilde{P}}_{\mathrm{kin}}| 0,\mathbf{E}_0
\rangle  \frac{\delta M}{M} = \langle
\mathbf{P}_{\mathrm{kin}}\rangle \frac{\delta M}{M}
\end{eqnarray}
Hence, the diverging contribution disappears entirely into the
inertial mass featuring in the kinetic momentum
${\mathbf{P}}_{\mathrm{kin}}= M\mathbf{v} $ of the atom.

We proceed with the converging terms. When we remove the
divergencies from the expression even the part with compensating
phase factors will yield a finite contribution. This can be shown to
be of order $\alpha^3 \mathbf{P}_A$ and will be ignored here. The
terms with exponentials typically have $k \approx 1/a_0$ since the
ground state has $r \approx a_0$. In that case $\hbar\omega_k \gg
e^2/4\pi \varepsilon_0 a_0$ and the atomic energy $H_0$ can be
neglected. Similarly the recoil energy $\hbar^2 k^2/2m_i$ is much
smaller than $\hbar \omega_k$. This simplification leads to the
following expression,
\begin{eqnarray}\label{ex2}
&&\langle\mathbf{P}_{\mathrm{long}}\rangle =\nonumber \\ && 2e^2
\sum_{\mathbf{k}} \frac{\mathcal{A}_k^2}{\hbar\omega_k}\left\{
\langle 0 ,\mathbf{E}_0|
 \mathbf{\Delta_k} \cdot \mathbf{p} \left(\frac{1}{m_2}e^{i\mathbf{k}\cdot \mathbf{r}} - \frac{1}{m_1} e^{-i\mathbf{k}\cdot \mathbf{r}} \right)
 \right. \nonumber \\ && - \mathbf{\Delta_k}\cdot \mathbf{Q}_0\frac{1}{M}
 \left(e^{i\mathbf{k}\cdot \mathbf{r}} + e^{-i\mathbf{k}\cdot \mathbf{r}}
 \right)\nonumber\\
  && + \left.\frac{e}{2}  \mathbf{\Delta_k}\cdot(\mathbf{B}_0\times \mathbf{r})
  \left(\frac{1}{m_2}e^{i\mathbf{k}\cdot \mathbf{r}} + \frac{1}{m_1} e^{-i\mathbf{k}\cdot \mathbf{r}} \right) \right\}|0,\mathbf{E}_0\rangle
\end{eqnarray}
with $ \mathbf{\Delta_k}= \sum_{\epsilon_k} \epsilon_k\epsilon_k$
the transverse projection matrix. The integral over $\mathbf{k}$ can
be carried out by using that $\int_0^\infty dk \int d\Omega
\exp(i\mathbf{k}\mathbf{\cdot r}) \mathbf{\Delta_k}=\pi^2
(1+\hat{\mathbf{r}}\hat{\mathbf{r}})/r$ and we get,

\begin{eqnarray}\label{ex2bis}
    && \langle\mathbf{P}_{\mathrm{long}}\rangle =\frac{e^2}{8\pi\varepsilon_0
    c_0^2} \langle 0,\mathbf{E}_0
    |\frac{1+\hat{\mathbf{r}}\hat{\mathbf{r}}}{r}\cdot \nonumber \\ && \left[ \mathbf{p}
    \left(\frac{1}{m_2}-\frac{1}{m_1}\right)
    -2\frac{\mathbf{Q}_0}{M}+ \frac{e}{2\mu }\mathbf{B}_0 \times
    \mathbf{r} \right]\left|0 ,\mathbf{E}_0\right\rangle
\end{eqnarray}
Since $\langle 0,\mathbf{Q}_0,\mathbf{E}_0 |-e^2/4\pi \varepsilon_0
r |0,\mathbf{Q}_0,\mathbf{E}_0\rangle = 2E_0 +
\mathcal{O}(\mathbf{E}_0^2)$, the middle term generates a
contribution,
\begin{equation}\label{longQ}
    \langle\mathbf{P}_{\mathrm{long}}\rangle_2=
    \frac{8}{3}\frac{E_0}{Mc_0^2}
\mathbf{Q}_0
\end{equation}
This can, in fact, be recognized as the contribution to the ground
state energy of the modified Coulomb field seen by the moving
charges, as expressed by the Darwin interaction \cite{LL}, and thus,
via the relativistic equivalence, to the inertial mass. The first
term in the expression for $
\langle\mathbf{P}_{\mathrm{long}}\rangle $  is more elaborate. We
can insert the expression (\ref{abquantum}) for the polarized ground
state $\left|0 ,\mathbf{E}_0\right\rangle$, and expand $\tilde{H}_0$
linearly into the external magnetic field. Some algebra leads to the
expression
\begin{eqnarray}\label{long1}
   && \langle\mathbf{P}_{\mathrm{long}}\rangle_1 =-\frac{e^4}{16\pi\varepsilon_0
    c_0^2} \frac{(m_2-m_1)^2}{m_1^2m_2^2}\epsilon_{nml}B_nE_k\nonumber \\ && \langle 0|
    \mathbf{p}\cdot\frac{1+\hat{\mathbf{r}}\hat{\mathbf{r}}}{r}
    \frac{1'}{(\tilde{H}_0-E_0)^2} r_mp_l r_k |0\rangle +c.c
\end{eqnarray}
This can be further simplified for the hydrogen isotropic 1S ground
state for which $p_i |0\rangle= (i\hbar/a_0) \hat{r}_i |0\rangle$,
and using that $m_1 \gg m_2 = m_e$ so that
\begin{equation}\label{long2}
    \langle\mathbf{P}_{\mathrm{long}}\rangle_{1} =- \mathbf{B}_0 \times \mathbf{E}_0   \frac{e^4\hbar^2}{4\pi\varepsilon_0
    c_0^2m_e^2a_0} \frac{1}{3} \langle 0|
     \frac{1}{r}\hat{r}_m
    \frac{1'}{(\tilde{H}_0-E_0)^2} r_m |0\rangle
\end{equation}
If we write  $\alpha(0)= 18\pi a_0^3$, this becomes

\begin{equation}\label{long3}
    \langle\mathbf{P}_{\mathrm{long}}\rangle_{1} =-\kappa_1 \alpha^2 \mathbf{P}_\mathrm{A}
    \end{equation}
   with $\mathbf{P}_\mathrm{A}= \varepsilon_0\alpha(0)\mathbf{B}_0 \times \mathbf{E}_0 $ and the dimensionless number
    \begin{equation} \label{kappa}
   \kappa_1= \frac{2}{27}\left( \frac{e^2}{4\pi\varepsilon_0 a_0}\right)^2 \sum_{n\neq 0 \ell m}
     \frac{\langle 0|
     r^{-1} \mathbf{\hat{r}}|n\ell m\rangle\cdot \langle n\ell m|\mathbf{r}
     |0\rangle}{(E_n-E_0)^2}
\end{equation}
Since $\kappa_1 > 0$,  this contribution \emph{lowers} the classical
Abraham momentum. The sum over all excited  states (with $\ell=1$
imposed by selection rule, the sum over $m$ equals 1) involves both
the bound states and the continuous spectrum. The sum over discrete
states can easily be done numerically using a recursion formula for
the hypergeometric functions \cite{hyperg}. With  $E_n
=-e^2/8\pi\varepsilon_0 a_0 n^2$ the numerical factor equals
$\kappa_2(D)=0.21$. The part associated with continuous spectrum is
much harder and we shall here assume that the continuous spectrum
consists of plane waves $\exp(i\mathbf{q\cdot r})/\sqrt{V}$ with
energy $E_\mathbf{q}= \hbar^2q^2/2m_e$, which is a usual
approximation in treatises of the photo-electric effect
\cite{loudon} and valid in principle only for $q> 1/a_0$. This leads
to ($y=qa_0$),
\begin{eqnarray}
&&\kappa_1(C)=\frac{8}{27}\sum_\mathbf{q}\frac{\langle 0 |
{r^{-1}\mathbf{\hat{r}}}{}|\mathbf{q}\rangle \cdot \langle\mathbf{
q} | \mathbf{r} | 0\rangle }{(a_0^2q^2 + 1)^2}=\nonumber
\\ &&\frac{8}{27}\frac{16}{\pi} \int_{y_{min}}^\infty dy
\frac{y^3}{(y^2+1)^3}\left( \frac{\arctan y}{y^2}
-\frac{1}{y\sqrt{y^2+1}}\right)
\end{eqnarray}
Choosing  $y_{min}=0$ or $y_{min}=1$ as lower limits gives $9.3
\cdot 10^{-3}
 < \kappa_1(C) < 1.4 \cdot 10^{-2}$. This is small compared to
$\kappa_1(D)$. We shall adopt $\kappa_1=0.22$ \cite{remark2}.

\bigskip

We finally evaluate the last term in Eq.~(\ref{ex2bis}). Upon
inserting the ground state~(\ref{abquantum}) perturbed by the
electric field. this leads to

\begin{equation}
    \langle\mathbf{P}_{\mathrm{long}}\rangle_{2}=\frac{e^2}{4\pi
    \varepsilon_0} \frac{e^2}{6m_ec_0^2} \mathbf{B}_0 \times \mathbf{E}_0
    \sum_j'
     \frac{\langle 0|
     \mathbf{\hat{r}}|j \rangle\cdot \langle j |\mathbf{r}
     |0\rangle}{E_j-E_0}
\end{equation}
We can write this as
\begin{equation}\label{long3bis}
    \langle\mathbf{P}_{\mathrm{long}}\rangle_{2} =+  \kappa_2 \alpha^2\mathbf{P}_\mathrm{A}
    \end{equation}
    with the dimensionless number
    \begin{equation} \label{kappa2}
   \kappa_2= \frac{1}{27} \frac{e^2}{4\pi\varepsilon_0 a_0^2} \sum_{n\neq 0 \ell=1 m}
     \frac{\langle 0|
      \mathbf{\hat{r}}|n\ell m\rangle\cdot \langle n\ell m|\mathbf{r}
     |0\rangle}{E_n-E_0}
\end{equation}
The same method as above yields $\kappa_2(D) =0.0796$ for the
discrete spectrum. For the continuous spectrum we find, again
assuming perfect plane waves,
\begin{equation}
\kappa_2(C)=\frac{256}{27\pi } \int_{y_{min}}^\infty dy
\frac{y^4}{(y^2+1)^6}\approx 0.018
\end{equation}
assuming a lower bound at $y=qa_0=1$. Thus $\kappa_2\approx  0.1$.

\bigskip

We see that  $\langle\mathbf{P}_{\mathrm{long}}\rangle_{1}$ en
$\langle\mathbf{P}_{\mathrm{long}}\rangle_{2}$ have opposite sign.
We conclude that  $\langle\mathbf{P}_{\mathrm{long}}\rangle \approx
-0.12 \alpha^2 \mathbf{P}_A$

\subsection{$
\langle\mathbf{P}_{\mathrm{trans}}\rangle $}

The leading contribution to the quantum expectation value of the
transverse momentum of electromagnetic field is given by,
\begin{eqnarray}\label{trans1}
 \langle \mathbf{P}_{\mathrm{trans}}\rangle& = & e^2  \sum_{\mathbf{k\epsilon}}  \mathcal{A}_k^2 \hbar\mathbf{ k} \times \nonumber \\
  && \langle 0,\mathbf{E}_0 |
 \epsilon\cdot \left(\frac{\mathbf{p}}{m_e} - \frac{\mathbf{Q}_0}{M}+\frac{e}{2m_e}
  \mathbf{B}_0\times \mathbf{r}\right) \nonumber \\ &&
\frac{1}{\left(
\tilde{H}_0(\mathbf{p}+\hbar\mathbf{k},\mathbf{Q}_0-\hbar\mathbf{k})-
  E_{0\mathbf{Q}_00}+ \hbar\omega_k\right)^2} \nonumber \\ && \epsilon\cdot \left(\frac{\mathbf{p}}{m_e} - \frac{\mathbf{Q}_0}{M}+\frac{e}{2m_e}
  \mathbf{B}_0\times \mathbf{r}\right) | 0 ,\mathbf{E}_0\rangle
   \nonumber
\end{eqnarray}
and a similar term for the proton (particle 1), which in fact can be
neglected for  $m_2 =m_e \ll m_1$. In leading order, atomic energies
can be neglected compared to typical photon energies $\hbar
\omega_k$. We need to expand the denominator into $\mathbf{k}$ in
order to find a non-vanishing contribution. This leads to,
\begin{eqnarray}\label{trans2}
\langle \mathbf{P}_{\mathrm{trans}}\rangle &=&
\frac{e^2\hbar}{2\varepsilon_0 c_0}  \int \frac{d^3\mathbf{k}
}{(2\pi)^3} \frac{2\hbar\mathbf{ k}}{k(\hbar^2k^2/2m_e +
\hbar\omega_k)^3} \nonumber \\ &&  \langle 0,\mathbf{E}_0 |
 \left(\frac{\mathbf{p}}{m_e} -
\frac{\mathbf{Q}_0}{M}+\frac{e}{2m_e}
  \mathbf{B}_0\times \mathbf{r}\right)\cdot \nonumber \\
&&    \mathbf{\Delta_k} \left[\frac{1}{m_e}\hbar \mathbf{k}\cdot
\frac{e}{2}(\mathbf{B}_0 \times \mathbf{r})+\frac{1}{m_e}\hbar
\mathbf{k}\cdot \mathbf{p}+\frac{1}{M}\hbar \mathbf{k}\cdot
\mathbf{Q}_0 \right] \nonumber \\ && \cdot
\left(\frac{\mathbf{p}}{m_e} - \frac{\mathbf{Q}_0}{M}+\frac{e}{2m_e}
  \mathbf{B}_0\times \mathbf{r}\right) |0,\mathbf{E}_0\rangle
  \nonumber
\end{eqnarray}
Since $\int_0^\infty dk k^3/(\hbar^2k^2/2m_e + \hbar\omega_k)^3
=m_e/\hbar^4c_0^2$ this can be evaluated. For instance, the typical
contribution linear to $\mathbf{Q}_0$ will be,
\begin{eqnarray*}
  \langle \mathbf{P}_{\mathrm{trans}}\rangle   &\simeq & \frac{e^2}{4\pi\varepsilon_0 } \frac{1}{\pi\hbar c_0^3 m_e}
  \langle 0| \mathbf{p}^2|0\rangle \frac{\mathbf{Q}_0 }{M}\\
   &\simeq&  \frac{e^2}{4\pi\varepsilon_0 } \frac{1}{\pi\hbar c_0^3
   m_e} m_e E_0 \frac{\mathbf{Q}_0 }{M} \nonumber \\
   &\simeq& \alpha \frac{E_0}{Mc_0^2} \mathbf{Q}_0
\end{eqnarray*}
Similarly we can estimate that terms proportional to
$\mathbf{B}_0\times \mathbf{r}$ generate a momentum of order
$\alpha^3 \mathbf{P}_\mathrm{A}$. Thus, $\langle
\mathbf{P}_{\mathrm{trans}}\rangle$ is typically a factor $\alpha$
smaller than the longitudinal momentum $\langle
\mathbf{P}_{\mathrm{long}}\rangle$ found in Eqs.~(\ref{longQ}) and
(\ref{long3}). It will be neglected in this work.

\section{Relativistic corrections to the Pseudo-momentum}

A particle with mass $m$ and kinetic momentum $\mathbf{p}$ achieves
a relativistic correction $-p^4/8c_0^2m^3$ to its kinetic energy.
When exposed to external electromagnetic fields , the particles
Hamiltonian achieves a term
\begin{eqnarray}\label{wrel}
     {H}_{\mathrm{rel}}= &-&\frac{1}{8c_0^2m_1^3}\left|\mathbf{p}+
    \frac{m_1}{M}\mathbf{P}-e\mathbf{A}_1\right|^4 \nonumber \\
     &-&\frac{1}{8c_0^2m_2^3}\left|-\mathbf{p}+
    \frac{m_2}{M}\mathbf{P}+e\mathbf{A}_2\right|^4
\end{eqnarray}
It is easily shown that  $\tilde{H}_{\mathrm{rel}}=
U^*{H}_{\mathrm{rel}}U$ commutes with the canonical total momentum
$\mathbf{P}$. The kinetic momentum is related to the conjugate total
momentum by the relation
$\mathbf{P}_{\mathrm{kin}}=M\mathbf{\dot{R}} =
M\partial_\mathbf{P}H.$ Equation (\ref{K}) for the pseudo-momentum
is still valid but Eq.~(\ref{KKK}) achieves an extra term
$\mathbf{P}_{\mathrm{rel}}= -M\partial_\mathbf{P}H_{\mathrm{rel}}$.
Straightforward algebra  leads to
\begin{eqnarray}\label{prel}
   \mathbf{\tilde{P}}_{\mathrm{rel}} =
   \frac{1}{2m_1^2c_0^2}&&\left(\mathbf{p}+\frac{m_1}{M}\mathbf{P}-\frac{e}{2}\mathbf{B}_0\times
   \mathbf{r}\right)^3 \nonumber \\
   + \frac{1}{2m_2^2c_0^2}&&\left(-\mathbf{p}+\frac{m_2}{M}\mathbf{P}-\frac{e}{2}\mathbf{B}_0\times
   \mathbf{r}\right)^3
\end{eqnarray}
with notation $\mathbf{w}^3 := (\mathbf{w\cdot w}) \mathbf{w}$. In
the following we collect different contributions to $\langle 0,
\mathbf{E}_0 |\mathbf{\tilde{P}}_{\mathrm{rel}} |0, \mathbf{E}_0
\rangle$, that are either linear in $\mathbf{Q}_0$ or in
$\mathbf{B}_0 \times \mathbf{E}_0$.

The contributions linear to $\mathbf{P}$ add up to
\begin{eqnarray}
   \langle \mathbf{P}_{\mathrm{rel}} \rangle_1 &=& \langle 0,
\mathbf{E}_0 | \frac{1}{2Mc_0^2} \frac{1}{\mu}  \left[\mathbf{p}^2
\mathbf{P} + 2 (\mathbf{p}\cdot \mathbf{P}) \mathbf{p}\right]
|0, \mathbf{E}_0 \rangle \nonumber \\
   &=& \frac{1}{M c_0^2} \frac{5}{3} \mathbf{Q}_0\langle 0,
\mathbf{E}_0 | \frac{\mathbf{p}^2}{2\mu}
|0, \mathbf{E}_0 \rangle  \nonumber \\
   &=& -\frac{5}{3}\frac{E_0}{Mc_0^2} \mathbf{Q}_0
\end{eqnarray}
 If we next
split off  $\mathbf{\tilde{P}}_{\mathrm{rel},1}$  from the equation
and assume that $m_1 \gg m_2=m_e$, we obtain,
\begin{eqnarray}
 \mathbf{\tilde{P}}_{\mathrm{rel}} &=& \mathbf{\tilde{P}}_{\mathrm{rel},1}
 -
\frac{1+\mathcal{O}(m_e^2/m_p^2)}{2m_e^2c_0^2}\left(\mathbf{p}+\frac{e}{2}\mathbf{B}_0\times
   \mathbf{r}\right)^3
\end{eqnarray}
 In Appendix B we show that

$$
\langle 0, \mathbf{E}_0 | \left(\mathbf{p} + \frac{e}{2}\frac{\Delta
m}{M}\mathbf{B}_0\times \mathbf{r}\right)^3 |0,\mathbf{E}_0\rangle=0
$$
Using this identity and a little algebra leads to,

\begin{eqnarray}
&&\langle \mathbf{P}_{\mathrm{rel}} \rangle = \langle
\mathbf{P}_{\mathrm{rel}} \rangle_1 -\frac{m_e}{M}
\frac{1}{2m_e^2c_0^2} \times  \nonumber \\ && \left(e^2\langle 0|
(\mathbf{E}_0\cdot \mathbf{r}) \frac{( \mathbf{B}_0\cdot
\mathbf{L})'}{\tilde{H}_0-E_0} \mathbf{p} |0\rangle +
\frac{e}{2}\langle 0,\mathbf{E}_0|\mathbf{p}^2 \mathbf{B}_0\times
\mathbf{r} | 0 ,\mathbf{E}_0\rangle \right)\nonumber
\end{eqnarray}
with $\mathbf{L}=\mathbf{r} \times \mathbf{p}$. This expression can
be evaluated for the hydrogen atom, but we  will here restrict to an
order of magnitude. It follows straightforwardly that
$(\cdots)/m_e^2c_0^2 \sim \alpha^2 \mathbf{P}_\mathrm{A}$. We thus
conclude that

\begin{eqnarray}
\langle \mathbf{P}_{\mathrm{rel}} \rangle
=-\frac{5}{3}\frac{E_0}{Mc_0^2} \mathbf{Q}_0 +
\mathcal{O}\left(\alpha^2\frac{m_e}{M}\right) \,
\varepsilon_0\alpha(0)\mathbf{B}_0\times \mathbf{E}_0
\end{eqnarray}
Both terms are  of order $\alpha^2 m_e/M$ and together with the QED
contribution (\ref{longQ}) the first adds up to the value
$E_0/Mc_0^2$ that we could have  anticipated  from the relativistic
equivalence principle that includes the bounding energy $E_0/c_0^2$
into inertial mass. This finding fits into the general notion that
Casimir energy has inertial
 mass that respects the equivalence principle \cite{jaekcel}.
 The relativistic correction to the Abraham momentum is a factor
$m_e/M$ smaller than the contribution from the quantum vacuum, found
in Eq.~(\ref{long3}).

There is also a modification of the static polarizability imposed by
special relativity.
 For the hydrogen atom the relative correction was calculated by Bartlett and
Power \cite{bartlett} to be $-\frac{28}{27}\alpha^2 $. This will
give a similar contribution to the Abraham momentum as the one
stemming from the quantum vacuum. We will here take the point of
view that the atom's relativistic correction  to the Abraham
momentum entering via the static polarizability is not counted as
Casimir momentum, and adopt $\mathbf{P}_A$ with either the\emph{
observed or the exactly calculated} $\alpha(0)$ as the true Abraham
momentum. This is also our experimental procedure \cite{geertprl12}.

\section{Conclusion}
In this work we have calculated the contribution of the quantum
vacuum to the Abraham force. Our approach treats the kinetic
momentum $M\mathbf{v}$ and the magneto-electric Abraham contribution
on equal footing. The conserved pseudo-momentum is the sum  of
kinetic and Abraham momentum, $\mathbf{K}= \mathbf{K}_{\mathrm{kin}}
+\mathbf{P}_\mathrm{A}$, so that the force is $\partial_t
\mathbf{K}_{\mathrm{kin}} =
M\mathbf{\ddot{R}}=-\partial_t\mathbf{P}_\mathrm{A}$. As expected
from the equivalence principle, our approach shows that the inertial
mass $M$ is affected by the binding energy of the atom, with a
significant contribution that can be viewed as stemming from the
longitudinal electromagnetic field of the quantum vacuum, but which
 in electrodynamics is better known as the Darwin interaction
associated with moving charges \cite{LL}. As for contribution of the
quantum vacuum to the Abraham momentum, our main result is that, at
least for the simple case of the hydrogen atom, divergencies can be
uniquely renormalized into the masses of electron and proton. The
finite small remainder is of relative order $-0.12\alpha^2 \pm
\mathcal{O}(\alpha^3)\sim 6 \cdot 10^{-6}$, thus reducing the
classical value, and stemming from the gauge potential $e\mathbf{A}$
of the quantum vacuum. The transverse virtual photons with momentum
$\hbar \mathbf{k}$ contribute only to order $\alpha^3$. The static
polarizability $\alpha(0)$ itself has a well-known relativistic
correction of order $-\alpha^2$, whereas the contribution  from the
quantum vacuum to $\alpha(0)$ is much smaller, only of order
$\alpha^2 m_e/M$. We have corrected an error in a previous
publication \cite{kawka} where a modification of order $\alpha$ was
predicted for the relative change in the Abraham momentum. In the
present work, we have also considered relativistic corrections to
the Abraham force and have concluded them to be factor $m_e/M = 5
\cdot 10^{-4} $ smaller than the contribution of the quantum vacuum.

The QED approach in this work, very likely to be valid albeit more
complex for more complex atoms and molecules,  solves the UV
catastrophe encountered in a semi-classical approach \cite{feigel}.
Casimir momentum exists, is finite, but is of relative order
$\alpha^2$. How such ``Casimir momentum" reveals itself in more
complex quantum systems is an important though difficult many-body
problem. One could speculate it to scale like $(Z\alpha)^2$, with
$Z$ the atomic number, as is known to be true for relativistic
corrections to the static polarizability of hydrogen-like atoms
\cite{bartlett}. If this is true the ''Casimir momentum" could
become within reach of experimental observation.
\bigskip

This work was supported by the ANR contract PHOTONIMPULS
ANR-09-BLAN-0088-01. We would like to thank Denis Basko and Thierry
Champel for useful help.

\appendix

\section{Appendix A}

To leading order, the normalization factor in Eq.~(\ref{w}) will
affect the expectation values of $\mathbf{P}$ or $\mathbf{B}_0
\times \mathbf{r}$. It is therefore sufficient to evaluate it for
$\mathbf{Q}_0=0$, $\mathbf{E}_0=0$ and $\mathbf{B}_0=0$ . If we also
assume that $m_1 \gg m_2 = m_e$ we can write
\begin{eqnarray}
    N &= &\frac{1}{2}\langle 0 0,0| W \frac{1'}{\left( \tilde{H}_0(\mathbf{p},\mathbf{Q},\mathbf{r}) + H_{\mathrm{F}}-E_0\right)^2} W | 0
    0,0\rangle\nonumber \\
    &=& \frac{1}{2}e^2 \sum_{\mathbf{k\epsilon}} \mathcal{A}_k^2 \, \, \langle0|
    \epsilon\cdot \mathbf{p}\frac{1}{m_e}e^{-i\mathbf{k\cdot
    r}}\nonumber \\ && \frac{1}{\left(\tilde{H}_0(\mathbf{p},\mathbf{Q}=\mathbf{k},\mathbf{r}) + \hbar
    \omega_k-E_0\right)^2} \epsilon\cdot \mathbf{p}\frac{1}{m_e}e^{i\mathbf{k\cdot
    r} }  |0\rangle \nonumber \\
&=& \frac{e^2}{2m_e^2} \sum_{\mathbf{k\epsilon}} \mathcal{A}_k^2 \,
\, \langle 0|
    \epsilon\cdot \mathbf{p} \nonumber \\ && \frac{1}{\left(\tilde{H}_0(\mathbf{p}+\hbar \mathbf{k},\mathbf{Q}=\mathbf{k},\mathbf{r}) + \hbar
    \omega_k-E_0\right)^2} \epsilon\cdot \mathbf{p}  |0\rangle
    \nonumber
\end{eqnarray}
If we neglect the $\mathbf{p}\cdot \hbar\mathbf{k}/m_e$ term in the
denominator, and use $\mathbf{p} |0\rangle= i(\hbar/a_0)
\mathbf{\hat{r}} |0\rangle$, one obtains
\begin{eqnarray}
    N &= & \frac{e^2}{2m_e^2}\frac{2}{3}\frac{\hbar^2}{a_0^2}\frac{\hbar}{2\varepsilon_0c_0}
    \nonumber \\ && \int
    \frac{d^3\mathbf{k}}{(2\pi)^3} \frac{1}{k} \sum_j'  \frac{ |\langle 0|\mathbf{\hat{r}}|j
    \rangle |^2}{(E_j-E_0+\hbar^2k^2/2m_e + \hbar\omega_k)^2}
    \nonumber
\end{eqnarray}
The $k$-integral has an effective lower limit at $k=(E_j-E_0)/\hbar
c_0$ below which $k$-dependence disappears and we estimate it as,
\begin{eqnarray}
   && \int_0^{\frac{E_j-E_0}{\hbar c_0}} dk \frac{k}{(E_j-E_0)^2}+  \int_{\frac{E_j-E_0}{\hbar
c_0}}^\infty dk \frac{k}{(\hbar^2k^2/2m_e + \hbar k c_0)^2}
\nonumber \\ &\approx &\frac{1}{(\hbar c_0)^2}\left(-\log
\frac{E_j-E_0}{\hbar c_0} -\frac{1}{2}\right) \nonumber
\end{eqnarray}
so that,
\begin{equation}
    N\approx \alpha^3 \frac{1}{\pi} \sum_{n=2}^\infty \left(-\log \frac{E_n-E_0}{\hbar c_0}
-\frac{1}{2}\right) |\langle 0|\mathbf{\hat{r}}|n,\ell=1 \rangle |^2
\nonumber
\end{equation}
Using a similar method to evaluate Eq.(\ref{kappa}), the sum over
Rydberg states at \emph{constant} logarithm can be evaluated to be
$0.336$. The typical value for the logarithm  used in calculations
of the Lamb shift is $-8.35$ \cite{milloni}. We conclude that $N
\approx 0.84 \alpha^3$. The continuous spectrum will have a small
additional contribution.

\section{Appendix B}

Let $\mathbf{w} \equiv \mathbf{p} + e(\Delta m/M)\mathbf{B}_0\times
\mathbf{r}/2$. We prove that $\langle 0,\mathbf{E}_0| \mathbf{w}^3
|0,\mathbf{E}_0\rangle=0$, with notation
$\mathbf{w}^3:=(\mathbf{w}\cdot \mathbf{w})\mathbf{w}$.  To this end
we notice that $\left[(\mathbf{w}^2)^2, \mathbf{r}\right] = i\hbar
\partial_\mathbf{p} (\mathbf{\mathbf{w}}^2)^2= 4i\hbar
\mathbf{w}^3 $ and that $\tilde{H}_0 = \mathbf{w}^2/2\mu +
F(\mathbf{Q}_0, \mathbf{r})$ in the subspace of the eigenvalue
$\mathbf{Q}_0$ of $\mathbf{P}$. We see that

\begin{eqnarray}
 \left(\frac{\mathbf{w}^2}{2\mu}\right)^2&=& (\tilde{H}_0)^2 -
F(\mathbf{Q}_0, \mathbf{r})^2
-\frac{\mathbf{w}^2}{2\mu}F(\mathbf{Q}_0,
\mathbf{r})-F(\mathbf{Q}_0,
\mathbf{r})\frac{\mathbf{w}^2}{2\mu}\nonumber \\ &=&(\tilde{H}_0)^2
-\tilde{H}_0F(\mathbf{Q}_0, \mathbf{r})-F(\mathbf{Q}_0,
\mathbf{r})\tilde{H}_0 + F(\mathbf{Q}_0, \mathbf{r})^2 \nonumber
\end{eqnarray}
Since $\left[F(\mathbf{Q}_0, \mathbf{r}),\mathbf{r}\right]=0$ and
$\tilde{H}_0 |0,\mathbf{E}_0\rangle = E_0 |0, \mathbf{E}_0\rangle $
it follows that

$$ \langle 0,\mathbf{E}_0| \mathbf{w}^3|0,\mathbf{E}_0\rangle = \frac{1}{4i\hbar}\langle
0,\mathbf{E}_0|\left[(\mathbf{w}^2)^2, \mathbf{r}\right]
|0,\mathbf{E}_0\rangle = 0
$$


\begin{thebibliography}{}
\bibitem{remark2}The same assumption of plane waves in the continuous spectrum can be adopted for the static polarizability
$\alpha(0)=4\pi \nu a_0^3$, for which it is known that $\nu=9/2$.
The discrete spectrum contributes $\nu_D= 3.66$. A lower limit $q=0$
would yield $\nu_C=2.32$ clearly much too large. The lower limit
$q=1/a_0$ yields $\nu_C=0.48$, too small. A reasonable choice is
somewhat lower than  $q=1/a_0$.
\end{thebibliography}
\end{document}